# CHARMONIUM PRODUCTION AT FNAL-E835 TRIPLET P STATES

N. PASTRONE
(on behalf of the E835 Collaboration)
*I.N.F.N., Via P. Giuria 1, 10125 Torino, Italy*

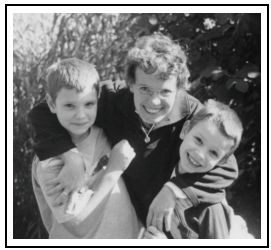

Fermilab experiment E835 has done an extensive set of studies on charmonium triplet P states. The most precise measurements on mass and total width come indeed from charmonium spectroscopy by direct formation of $c\bar{c}$ states in $\bar{p}p$ annihilations. Results on both partial widths and higher multipoles radiative transitions $\chi_{cJ} \to J/\psi\gamma \to \gamma e^+e^-$ are given, together with the $\gamma\gamma$ widths of $\chi_{c0}$ and $\chi_{c2}$. For the first time in hadronic charmonium production, hadronic annihilation channels ($\pi^0\pi^0$ and $\eta\eta$) have been clearly identified at the $\chi_{c0}$ resonance.

## 1 Introduction

Precision measurements of the ($c\bar{c}$) system (masses, widths, energy splittings, decay rates and their ratios) are important inputs to test the limit of pQCD and the order of magnitude of relativistic and radiative corrections.

In $\bar{p}p$ annihilations, where all $J^{PC}$ states can be formed via 2 or 3 perturbative gluons, P-wave charmonium states are directly accessible.

The experimental technique, pioneered at CERN by experiment R704 in 1983-84 [1], was fully exploited by experiment E760 at the Fermilab Antiproton Accumulator (1989-91) which gave the most accurate measurements available so far on mass and total width of $\chi_{c1}$ and $\chi_{c2}$ [2].

FNAL experiment E835, with a major upgrade of E760 apparatus, collected data in two different periods: 143 pb$^{-1}$ in Run I ('96-'97) and 113 pb$^{-1}$ in Run II (2000).

Charmonium resonances are scanned using the tunable, stochastically cooled $\bar{p}$ beam that intersects a hydrogen gas jet target. Instantaneous luminosity can be kept almost constant ($L \sim 2 \times 10^{31}$cm$^{-2}$s$^{-1}$) by increasing the jet target density ($\rho_{max} \sim 3.0 \times 10^{14}$ atoms/cm$^3$) as the circulating antiproton current decreases throughout each stack. At each energy point during a scan, the cross section is measured by normalizing the number of events which satisfy the event selection criteria for the final state under study to the integrated luminosity collected. Typically, an excitation curve is extracted from the huge hadronic background ($\frac{\sigma(\bar{p}p \to (c\bar{c})_R)}{\sigma(\bar{p}p \to X)} \leq 10^{-5}$) by tagging electromagnetic final states. Hence, resonance parameters may be determined without relying on the detector resolution, but only on the event statistics and on the knowledge of the $\bar{p}$ beam. At any setting of the beam momentum, the $\bar{p}p$ center of mass energy ($E_{cm}$) is known to 0.2 MeV; the spread (r.m.s.) in $E_{cm}$ is on average about 0.4 MeV. This resolution is substantially better than could be provided by detection equipment alone.

The non magnetic spectrometer was optimized for detection and identification of photons and electrons. It consists of a cylindrical central barrel and a planar forward system with full azimuthal coverage and polar angle acceptance from $2°$ to $70°$.

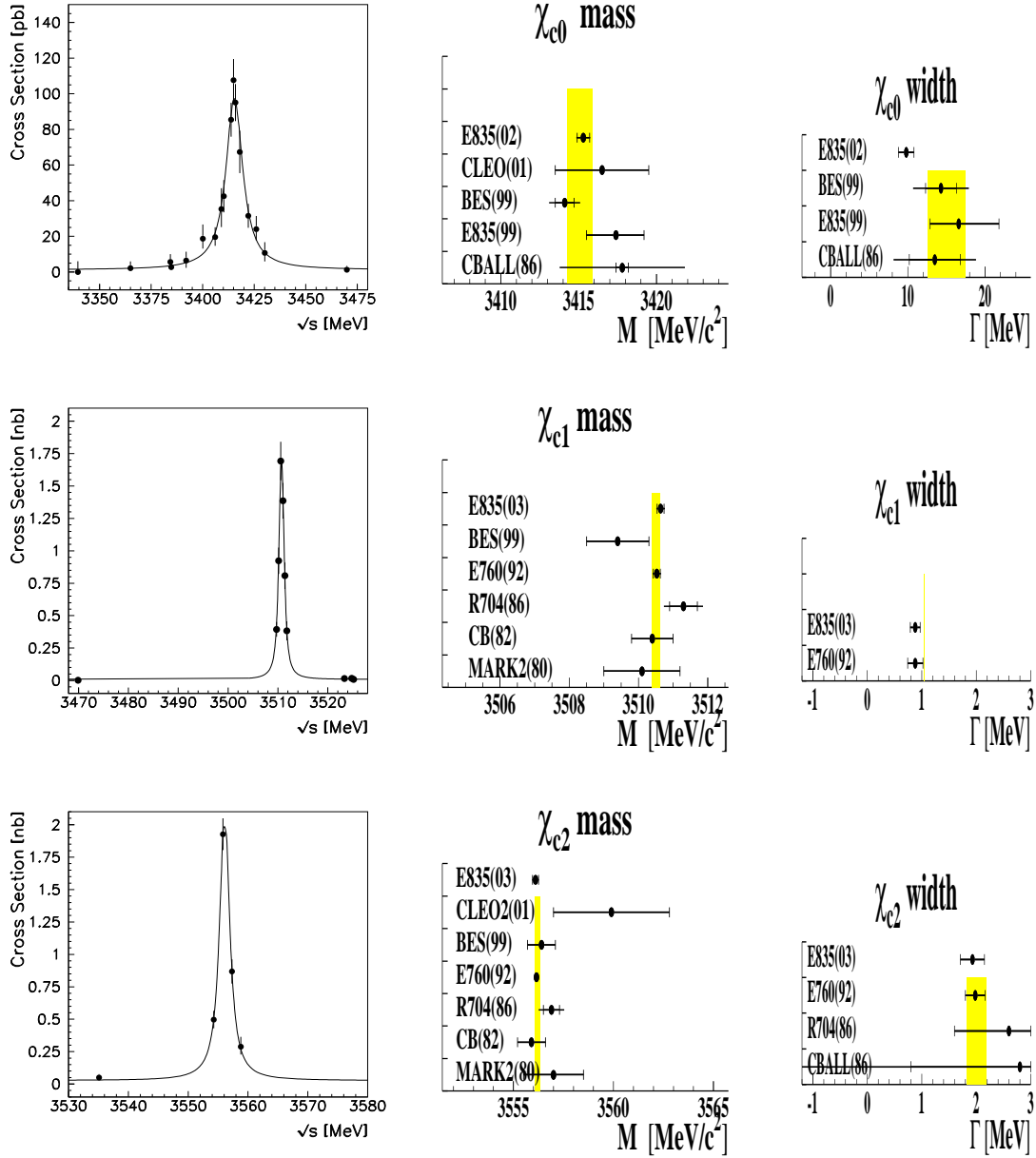

Figure 1: E835 $\chi_{cJ}$ states excitation curves, obtained selecting $J/\psi\gamma \to e^+e^-\gamma$ final states. A comparison of mass and total width measurements is also shown.

During Run I, E835 collected large samples of $\chi_{c1}$ and $\chi_{c2}$ events, mostly on the resonance peaks, to study the angular distribution of their radiative decay, as well as $\chi_{c2} \to \gamma\gamma$ partial width.

A major upgrade of the Accumulator (1997-2000) allowed a smooth running at the $\chi_{c0}$ region in Run II, while a new measurement of $\chi_{c1}$ and $\chi_{c2}$ parameters proves the consistency of the results and the understanding of related systematic uncertainties.

For the first time the $\chi_{c0}$ production in $p\bar{p}$ annihilations has been also detected selecting hadronic final states ($\pi^0\pi^0$ and $\eta\eta$).

## 2  $p\bar{p} \to \chi_{cJ} \to \gamma J/\psi \to \gamma(e^+e^-)$.

A new set of measurements of mass, total width and product of B($\chi_{cJ} \to \bar{p}p$) $\times \Gamma(\chi_{cJ} \to J/\psi\,\gamma)$ comes from E835 Run II.

On a clean identified sample of $e^+$ $e^-$ with high invariant mass, a 5C kinematic fit is applied to select $p\bar{p} \to \chi_{cJ} \to \gamma J/\psi \to \gamma(e^+e^-)$ events. $\chi_{cJ}$ parameters are directly measured by a maximum likelihood fit of the excitation curve (see Fig. 1) with a Breit-Wigner resonant cross section (convoluted with the beam energy distribution) and a flat background cross section.

A total luminosity of $\sim 32.8$ nb$^{-1}$ (a factor 6 larger than in Run I) was devoted in year 2000 [3] to measure the $\chi_{c0}$ resonance parameters, not studied by E760. Results are summarized in Table 1 together with the results on $\chi_{c1}$ and $\chi_{c2}$ both from E835 (Run II) and E760, that show the high stability of the measurements, after 10 years and a completely modified $\bar{p}$ source.

From the full E835 sample of $\sim 1.0$ M $\psi\prime$ events (same sample as Crystal Ball) the analysis of $\sim 2000$ events $\psi\prime \to J/\psi\gamma\gamma$ will give access to the branching ratios to $\gamma\chi_c$, $\psi\pi^0$, $\psi\eta$ final states. Analysis is under way to study B($\psi\prime \to \gamma\chi_{cJ}$)$\times$ B($\chi_{cJ} \to J/\psi\gamma$) as a complement of $\chi_{cJ}$ measurements.

The Run I data sample of $\sim 6000$ $\chi_{c2}$ and $\sim 2100$ $\chi_{c1}$ allowed the study of angular distribution [4] of the process $p\bar{p} \to \chi_{c1,2} \to J/\psi\gamma \to e^+e^-\gamma$ to measure deviations from pure E1 transition in E1-M2 and E1-E3 interference terms. Relativistic effects are expected to cancel in the fractional amplitude $a_2 \sim$ M2/E1.

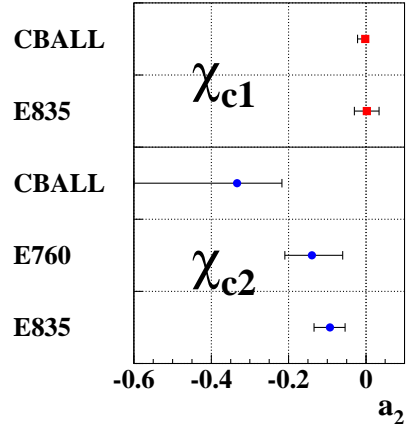

Figure 2: $a_2 \sim$ M2/E1 results

From a relativistic calculation [5], it is expected a ratio $a_2(\chi_{c1})/a_2(\chi_{c2})\sim 0.68$ while E835 shows a $2\sigma$ discrepancy (see results in Fig. 2, compared with previous experiments).

The helicity 0 contribution to $\Gamma(\chi_{c2} \to \bar{p}p)$ was also measured: $B_0^2$=13$\pm$8.

## 3  $\chi_{c0,2} \to \gamma\gamma$

E835 has published [6] a study of $\gamma\gamma$ decays of $\chi_{c0}$ and $\chi_{c2}$ from Run I. $\gamma\gamma$ braching ratio can be determined from the ratio: $R_{\gamma\gamma} = B(\chi_{c0,2} \to \gamma\gamma)/B(\chi_{c0,2} \to J\psi\gamma \to e^+e^-\gamma)$
where the systematic errors on beam momentum and luminosity cancel since rates are measured simultaneously and the error is not affected by the relatively well measured B($\chi_{c0,2} \to J\psi\gamma$).

$R_{\gamma\gamma}$ was measured to be (0.244$\pm$0.125) and (1.67$\pm$0.30)$\times$ 10$^{-2}$ respectively at $\chi_{c0}$ and $\chi_{c2}$.

Since the background, mainly due to $\pi^0\pi^0$ and $\pi^0\gamma$ events with one or two undetected photons, has a strongly forward-peaked angular distribution, events are selected using an acceptance cut to maximize signal to background ratio ($<$ 1). $\pi^0\pi^0$ and $\pi^0\gamma$ cross sections are determined from the data, while a fast calorimeter simulation is used to estimate the number of events contributing to the $\gamma\gamma$ background.

Fig. 3 shows the $\gamma\gamma$ data used in the $\chi_{c2}$ analysis with an acceptance cut $cos\theta^* < 0.45$.

The latest results on $\gamma\gamma$ partial width ($\Gamma_{\gamma\gamma}$) are compared in Fig. 3.

Table 1: Final values from Fermilab Charmonium Spectroscopy experiments.

| $\chi_{c0}$ | | |
|---|---|---|
| PARAMETER | E835 | E760 |
| M [MeV/c$^2$] | 3515.5 ± 0.4 ± 0.2 | – |
| $\Gamma_{TOT}$ [MeV] | 10.1 ± 1.0 | – |
| $B_{in} \times \Gamma(\chi_{c0} \to J/\psi + \gamma)$[eV] | 27.9 ± 2.5 ± 0.7 | – |
| $\chi_{c1}$ | | |
| PARAMETER | E835 | E760 |
| M [MeV/c$^2$] | 3510.64 ± 0.10 ± 0.07 | 3510.53 ± 0.10 ± 0.07 |
| $\Gamma_{TOT}$ [MeV] | 0.88 ± 0.09 | 0.88 ± 0.14 |
| $B_{in} \times \Gamma(\chi_{c1} \to J/\psi + \gamma)$[eV] | 18.8 ± 0.7 ± 0.6 | 21.8 ± 2.7 ± 1.2 |
| $\chi_{c2}$ | | |
| PARAMETER | E835 | E760 |
| M [MeV/c$^2$] | 3556.10 ± 0.15 ± 0.07 | 3556.15 ± 0.11 ± 0.07 |
| $\Gamma_{TOT}$ [MeV] | 1.93 ± 0.22 | 1.98 ± 0.18 |
| $B_{in} \times \Gamma(\chi_{c2} \to J/\psi + \gamma)$[eV] | 25.8 ± 1.9 ± 0.8 | 28.2 ± 2.9 ± 1.5 |

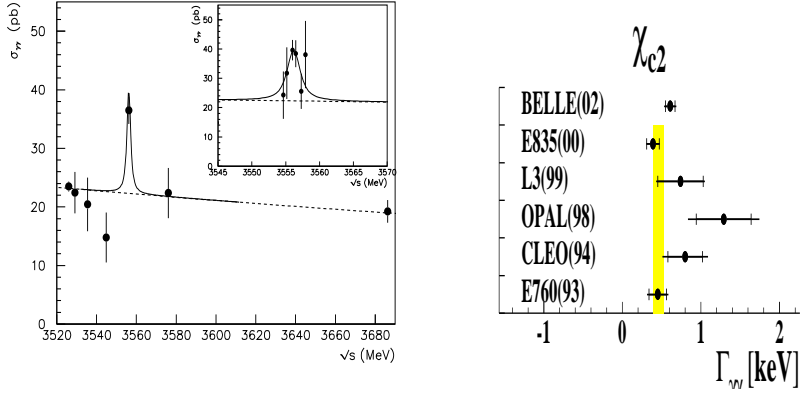

Figure 3: Measured $\gamma\gamma$ cross section at $\chi_{c2}$: the solid curve is the fit of signal+background with fix M and $\Gamma$. A comparison of $\Gamma_{\gamma\gamma}(\chi_{c2})$ results is shown on the right.

The higher statistics collected in Run II allowed a new measurement of the decay of the $\chi_{c0}$ to two photons.

The incoherent $\gamma\gamma$ background from $\pi^0\pi^0$ and $\pi^0\gamma$ (feeddown) is estimated by MC at each energy point and suggests a cut at $|\cos\theta^*| < 0.4$. However, coherent continuum $\bar{p}p \to \gamma\gamma$ ($\sim$ 10 pb up to $|\cos\theta^*| = 0.6$, according to CLEO[7] and VENUS[8]) can interfere with $\gamma\gamma$ resonant production and reduce the optimal angular region. Fig. 4 shows a clear interference for $|\cos\theta^*| > 0.2$. To determine the $\chi_{c0}$ branching ratio to two photons, the measured cross section is fit to a Breit-Wigner resonance and background from feeddown, plus an interference term to account for $\bar{p}p \to \gamma\gamma$ continuum. The mass and width are fixed to values obtained in $J/\psi\gamma$ decay [3].
The results is: B($\chi_{c0} \to p\bar{p}$)$\times$ B($\chi_{c0} \to \gamma\gamma$) = (6.52 ± 1.18(sta) ± 0.55(sys))$\times 10^{-8}$.
Using B($\chi_{c0} \to p\bar{p}$) = (2.2±0.5)$\times 10^{-4}$ from PDG2002 [9] and $\Gamma_{\chi_{c0}}^{E835}$=9.8±1.0 MeV we obtain:
$\Gamma_{\gamma\gamma}$ = (2.90 ± 0.59(sta+sys) ± 0.66(BR) ± 0.3($\Gamma$)) keV.
A comparison of experimenthal results is shown in Fig. 4 (right).

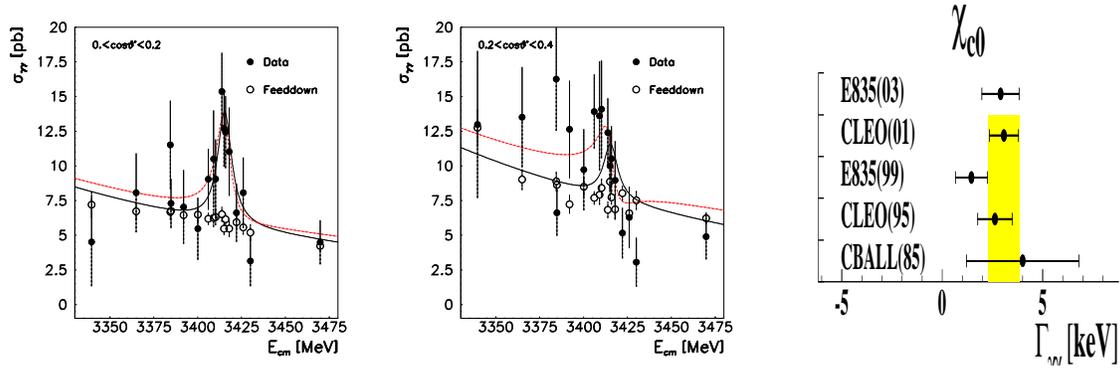

Figure 4: $\gamma\gamma$ cross section in two angular region ($0. < \cos\theta^* < 0.2$ and $0.2 < \cos\theta^* < 0.4$): the best fit results with (dashed line) and without (solid line) interference are included. A summary of the $\Gamma_{\gamma\gamma}(\chi_{c0})$ results is shown.

## 4  $\bar{p}p \to \chi_{c0} \to \pi^0\pi^0 \ (\eta\eta)$

E835-II observed the decay into $\pi^0\pi^0$ and $\eta\eta$ of the $\chi_{c0}$ state produced in $\bar{p}p$ annihilations. Although the resonant amplitude is an order of magnitude smaller than that of the non-resonant continuum, interference between resonant and continuum gives a sizeable signal at $\cos\theta^* \sim 0$. ($\theta^*$ is the $\pi^0\pi^0$ ($\eta\eta$) production angle in the center of mass frame with respect to the $\bar{p}$ direction). In the vicinity of the $\chi_{c0}$, the differential cross section $d\sigma/d|\cos\theta^*|$ of the $\bar{p}p \to \pi^0\pi^0$ reaction can be written as:

$$\frac{d\sigma}{dz}(x,z) = \left|\frac{-A_R}{x+i} + A_I \, e^{i\delta_I}\right|^2 + \left|A_N \, e^{i\delta_N}\right|^2 = \frac{A_R^2}{x^2+1} + A_I^2 + 2\, A_R\, A_I \, \frac{\sin\delta_I - x\cos\delta_I}{x^2+1} + A_N^2$$

where $x = \frac{E_{CM} - M_{\chi_0}}{\Gamma_{\chi_0}/2}$ and $z \equiv |\cos\theta^*|$.

The initial $\bar{p}p$ state has two ortogonal helicity states $\lambda = |\lambda_{\bar{p}} - \lambda_p| = 0, 1$, not interfering with one other. The $\chi_{c0}$ is only produced in $\lambda=0$ initial state. $A_R$ is the Resonant Amplitude, $A_I$ the Interfering part of the Non-Resonant A amplitude ($\lambda=0$), and $A_N$ the Non-Interfering part of the Non-Resonant Amplitude ($\lambda=1$).

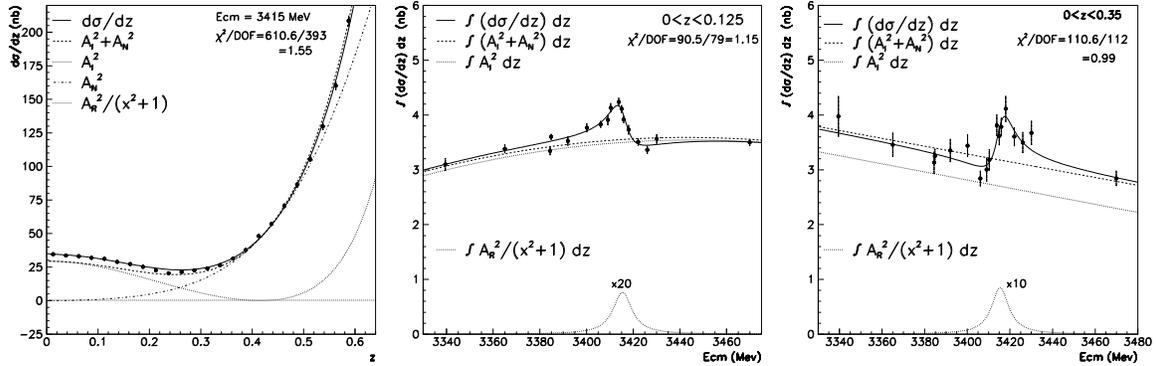

Figure 5: Measured $\pi^0\pi^0$ cross section versus $z \equiv |\cos\theta^*|$ at $E_{cm}= 3415$ MeV(left) and versus $E_{cm}$ integrated over $0<z<0.125$ (center). Measured $\eta\eta$ cross section versus $E_{cm}$ integrated over $0<z<0.35$ (right). Expected contribution from the pure resonance is shown, magnified, at the bottom (center and right).

Fig. 5(left) shows the several components of the differential cross section. The effect of the resonance, amplified by the interference, is seen in the gap (evident only at small z) between $d\sigma/dz$ and $A_I^2 + A_N^2$. The measured cross section, over a reduced z range, appears clearly enhanced and re-shaped by the interference (see Fig. 5, center and right).

The product of the input and output branching fractions is extracted [11] from a partial wave expansion fit: $B(\chi_{c0} \to p\bar{p}) \times B(\chi_{c0} \to \pi^0\pi^0) = (5.09 \pm 0.81(\text{sta}) \pm 0.25(\text{sys})) \times 10^{-7}$.

With the same technique, the $\eta\eta$ final state, with about a factor 20 lower statistics, was also analyzed [12] to determine: $B(\chi_{c0} \to p\bar{p}) \times B(\chi_{c0} \to \eta\eta) = (4.0 \pm 1.2_{stat}{}^{+0.5}_{-0.3sys}) \times 10^{-7}$.

No resonant signal was observed into the isospin-suppressed $\pi^0\eta$ final state, used as a control channel to check the systematics of the experiment.

## 5  Conclusions & perspectives

Study of charmonium P states in $\bar{p}p$ has shown to be very powerful.

A new generation of $\bar{p}p$ experiments can reduce the current statistical and systematic errors to provide the most precise measurements of $M$ and $\Gamma_{TOT}$ for $\chi_{cJ}$ states.

$\gamma\gamma$ decays are complementary to production in two photon collisions at $e^+e^-$ machines.

It was pointed out [13] that determination of quantity like $\Gamma(\chi_{c2} \to \gamma\gamma)$ depends directly from two processes: $\bar{p}p \to \chi_{c2} \to \gamma\gamma$ and $\gamma\gamma \to \chi_{c2} \to J/\psi\gamma, hadrons$ and indirectly from other quantities deduced from: $\psi\prime \to \gamma\chi_c \to J/\psi\gamma\gamma$, $\psi\prime \to J/\psi\pi\pi$, and $\bar{p}p \to \chi_{c2} \to J/\psi\gamma$.

Recently a global refitting of all existing data on charmonium that uses as inputs measured combinations of branching fractions and partial widths has been implemented in an attempt to resolve the issue of correlations in the derivation of individual parameter.

Global refitting in PDG 2002 [9] of all $\chi_c$ and $\psi\prime$ data gives substantial changes in the overall branching ratios and radiative widths (i. e. $B(\chi_{c2} \to J/\psi\gamma)$ raised from 13.6% to 18.7%).

This global refitting reduces discrepancies between $\bar{p}p$ and $e^+e^-$ experiments on $\Gamma(\gamma\gamma)$ and $\Gamma(\bar{p}p)$, but to fully exploit the accuracy of $\bar{p}p$ measurements on products of branching ratios, new $e^+e^-$ data are still needed.

The first $c\bar{c}$ signal in pure hadronic channels ($\bar{p}p \to \pi^0\pi^0, \eta\eta$) was observed, exploiting the interference in scattering at 90° in the CM. The effectiveness of this technique, to extract a resonant signal in channels dominated by order-of-magnitude larger non-resonant cross sections, looks very promising to study elusive states as siglets $c\bar{c}$ or hadromolecules $c\bar{c}q\bar{q}$.

## References

1. C. Baglin et al., *Phys. Lett.* B **172**, 455 (1986).
2. T. A. Armstrong et al., *Nucl. Phys.* B **373**, 35 (1992).
3. S. Bagnasco et al., *Phys. Lett.* B **533**, 237 (2002).
4. M. Ambrogiani et al., *Phys. Rev.* D **65**, 052002 (2002).
5. K.J. Sebastian, H. Grotch and F.L. Ridener, *Phys. Rev.* D **45**, 3163 (1992).
6. M. Ambrogiani et al., *Phys. Rev.* D **62**, 052002 (2000).
7. M. Artuso et al., *Phys. Rev.* D **50**, 5084 (1994).
8. H. Hamasaki et al., *Phys. Lett.* B **407**, 185 (1997).
9. K.Hagivara et al. (Particles Data Group), *Phys. Rev.* D **66**, 010001 (2002).
10. M. Graham, Ph. D. Thesis, University of Minnesota, 2002, Fermilab-thesis-2002-19.
11. M. Andreotti et al., submitted to *Phys. Rev. Lett.*
12. P. Rumerio, Ph. D. Thesis, Northwestern University, Evanston, Fermilab-thesis-2003-04.
13. C. Patrignani, *Phys. Rev.* D **64**, 034017 (2001).